\documentclass[12pt,preprint]{aastex}
\usepackage{natbib}
\usepackage{graphicx}
\usepackage{wasysym}
\usepackage{txfonts}
\def\fH2{\mbox {f$_{{\rm H}_2}$}}

\def\EBV{\mbox{E$_{\rm B-V}$}}

\def\nH2{\mbox{${\rm n}(\HH$)}}
\def\enH2{\mbox{$n_{(\HH$)}}}

\def\pccc{~{\rm cm}^{-3}} 
 
\def\pcc {\mbox{${~{\rm cm}^{-2}}$}}

\def\Tsub#1 {\mbox{${\rm T}_{\rm #1}$}}
\def\TK  {\Tsub K }

\def\degr{$^{\rm o}$}
\def\p{\mbox{$^+$}}

\def\cch{\mbox{C$_2$H}}
\def\cfh{\mbox{C$_4$H}} 
\def\anicfh{\mbox{C$_4$H$^-$}}

\def\h13cop{\mbox{{H$^{13}$CO\p}}}

\def\c3h2{\mbox{C$_3$H$_2$}}
\def\linear{\mbox{$l$-C$_3$H$_2$}}
\def\cyclic{\mbox{$c$-C$_3$H$_2$}}
 
 \def\R0{R$_0$}
\def\G0{\mbox{G$_0$}}

\def\ddeg{{}^\circ\kern-.1em}

\def\kms{\mbox{km\,s$^{-1}$}}
\def\ps{\mbox{s$^{-1}$}}

\def\E#1 {$10^{#1}$}
\def\E#1 {E{#1}}
\def\P#1,{$\nH2\TK~=~#1\times~10^4\pccc$~K}
\def\ec#1,#2,#3,{#1\,(#2)\E{#3}}

\def\H3{\mbox{H$_3$}}

\def\RH2{\mbox{R$_{\rm G}$}}
\def\g13{\mbox{g$_{13}$}}

\def\kHeH2{\mbox{$k_{ He-\HH}$}}

\def\tim#1,#2{\mbox{{$#1\times10^{#2}$}}}

\newcommand{\emm}[1]{\ensuremath{#1}}   
\newcommand{\emr}[1]{\emm{\mathrm{#1}}} 

\newcommand{\HH}{\emr{H_2}}

\slugcomment{generated \today}

\shorttitle{Not really, John}
\shortauthors{Liszt, Sonnentrucker, Cordiner & Gerin}

\begin{document}


\title{The abundance of C$_3$H$_2$ and other small hydrocarbons in the diffuse 
interstellar  medium
\thanks{Based on observations obtained with the NRAO Jansky Very Large Array (VLA).}}


\author{Harvey Liszt}
\affil{National Radio Astronomy Observatory \\
        520 Edgemont Road, Charlottesville, VA 22903-2475}
\and

\author{Paule Sonnentrucker}
\affil{Space Telescope Science Institute \\
3700 San Martin Dr, Baltimore, MD 21218}

\and

\author{Martin Cordiner}
\affil{Astrochemistry Laboratory and the Goddard Center for Astrobiology \\
  Mailstop 691, NASA Goddard Space Flight Center \\
 8800 Greenbelt Road, Greenbelt, MD 20770, USA}

\and

\author{Maryvonne Gerin}
\affil{LERMA, UMR 8112 du CNRS, Observatoire de Paris \\
  \'Ecole Normale Sup\'erieure, UPMC \& UCP, France}

\email{hliszt@nrao.edu}




\begin{abstract}
Hydrocarbons are ubiquitous in the interstellar medium, observed in diverse 
environments ranging from diffuse to molecular dark clouds and strong 
photon-dominated regions near HII regions.  Recently, two broad diffuse 
interstellar bands (DIBs) at  4881\AA\ and 5450\AA\ were attributed to the 
linear version of propynylidene \linear, a species whose more stable cyclic 
conformer \cyclic\
has been widely observed in the diffuse interstellar medium at radio wavelengths.  
This attribution has already been criticized on the basis of indirect 
plausibility arguments because the required column densities are quite 
large, N(\linear)/\EBV\ $= 4 \times 10^{14}~\pcc$ mag$^{-1}$.  Here 
we present new measurements of N(\linear) based on simultaneous 18-21 
GHz VLA absorption profiles of cyclic and linear  C$_3$H$_2$taken along 
sightlines toward extragalactic radiocontinuum background sources with 
foreground Galactic reddening \EBV\ = 0.1 - 1.6 mag.  We find that 
N(\linear)/N(\cyclic) $\approx 1/15 - 1/40$ and 
N(\linear)/\EBV\ $\approx 2\pm 1 \times10^{11}\pcc$ mag$^{-1}$, so that the 
column densities of \linear\ needed to explain the diffuse interstellar 
bands are some three orders of magnitude higher than what is observed.  
We also find N(\cfh)/\EBV\ $<1.3 \times10^{13}\pcc$ mag$^{-1}$ 
and N(\anicfh)/\EBV\ $< 1\times 10^{11}\pcc$ mag$^{-1}$ ($3\sigma$).
Using available data for CH and \cch\ we compare the abundances 
of small hydrocarbons in diffuse and dark clouds as a guide to their 
ability to contribute as DIB carriers over a wide range of conditions in the
interstellar medium.

\end{abstract}



\section{Introduction}

The cyclic (ring) conformer of propynylidene, \cyclic, was discovered and 
identified 
in the interstellar medium (ISM) by \cite{MatIrv85} and \cite{ThaVrt85} and 
quickly recognized as a ubiquitous tracer of molecular gas in both dark 
\citep{MadIrv+89,CoxWal+89} and diffuse clouds \citep{CoxGue+88}.  It is 
abundant even in strong photon-dominated regions (PDR) like the Horsehead 
Nebula \citep{PetTey+05}.  Extensive recent surveys of the small hydrocarbons 
\cyclic\ and \cch\ in mm-wave absorption \citep{LucLis00C2H,GerKaz+11} show that 
they have a nearly-fixed relative abundance N(\cch)/N(\cyclic) $\approx$ 20 in 
diffuse gas, much above that seen in the dark cloud TMC-1 where N(\cch)/N(\cyclic)
$\approx 2$ \citep{OhiIrv+92}, and a nearly constant abundance with respect to 
\HH, X(\cyclic) = N(\cyclic)/N(\HH) $\approx 2-3\times 10^{-9}$.

By contrast, when the less stable linear conformer \linear\ was detected in the 
ISM \citep{CerGot+91} it was found to be much less abundant
than \cyclic\ in the dark cloud TMC1, with N(\cyclic)/N(\linear) $\approx$ 100
and N(\linear) = $2.5\times10^{12}\pcc$.    Subsequent observations of 
 \linear\ in mm-wave absorption from 
diffuse clouds in front of the distant HII regions W49 and W51 
\citep{CerCox+99} found that the ratio \cyclic/\linear\ was smaller,
N(\cyclic)/N(\linear) = 3-7,  although with small column densities overall, 
N(\linear) $\la 5\times10^{11}\pcc$.

Despite strong suggestions that column densities of \linear\ are modest in the
ISM, whether in diffuse or dark gas, \cite{MaiWal+11} recently
hypothesized that broad diffuse interstellar bands (DIBs) at 4881\AA\ and 5450\AA\ 
are produced by a ground-state electronic transition of \linear, with N(\linear) 
$\approx 2-5\times10^{14}$ toward HD 206267 (\EBV\ = 0.52 mag) and HD 183143 
(\EBV\ = 1.27 mag).  This assignment has already been criticized for obvious 
reasons based on indirect plausibility arguments \citep{KreGal+11,OkaMcC11}.  
Here we show directly  that the diffuse ISM is in no way anomalous with regard 
to the \linear\ abundance.  Seen in absorption originating in Galactic 
diffuse and translucent clouds occulting 4 compact extragalactic 
mm-wave continuum sources along 
sightlines with \EBV\ = 0.1 mag - 1.6 mag, column densities of \linear\
are indeed small, with N(\linear) $\la 1 - 3\times10^{11}\pcc$.

The plan of this work is as follows.  In Section 2 we describe new
18-21 GHz absorption line observations of \linear, \cyclic, \cfh\ and 
\anicfh.  In Section 3 we discuss the abundances and column densities
of these and other small hydrocarbons so that their ability to contribute
as carriers of DIBs may be accurately assessed.

\section{Observations, conventions and conversion from optical depth to column density}

The new observations reported here were taken at the National Radio
Observatory's Jansky Very Large Array (VLA) on 16 and 17 September 2011 under
proposals 11B/003 (for  C$_3$H$_2$) and 11B/076 (for \cfh\ and its anion \anicfh),
in  ``move'' time during the A-D configuration transition.  
The data were taken in four schedule blocks (SB) of 2 hour duration
observing in two orthogonal polarizations in each of two basebands
with all of the 4 basebands covered by 6 contiguous spectral windows
having 128 independent channels over a bandwidth 2 MHz.  The spectral lines  
observed here, all in the range 18-21 GHz,  are summarized in Table 1.
At 20 GHz, a bandwidth of 2 MHz corresponds to 30 \kms\ so that the widths of 
the spectral line channels were all approximately 30 \kms/128 $\approx$ 
0.25 \kms.  All velocities discussed here are relative to the kinematic
definition of the Local Standard of Rest that is in universal use at
radio telescopes.

Properties of the sources and sightlines observed and their integrated
molecular optical depths are summarized in Table 2; 
two targets and a bandpass calibrator (3C84) were covered in each SB.
Considerable time was devoted to reference pointing on each continuum source 
before it was observed.  No absolute amplitude calibration was performed but 
the fluxes relative to that of the bandbass calibrator 
(S$_\nu \approx 16$ Jy) are given in Table 2. In each SB the bandpass calibrator 
was observed for approximately 20m.  B2251+158 was observed for approximately 
18 minutes and the other sources for approximately 40m during any one SB.  
The weather and system temperatures were significantly better during the observing 
for 11B/003 (\linear\ and \cyclic) as reflected in the rms noise levels reported in 
Table 2. As well, some unexplained malfunction resulted in loss of much of the flux 
during the search for \cfh\ toward 3C111, as reflected in Table 2 and beyond.

The data were reduced using very standard techniques in CASA.   The bandpass 
calibrator observations were phase-calibrated during each scan sub-interval 
followed by construction of an average bandpass solution.  This was applied on 
the fly to complex gain-cal solutions for each continuum target at the sub-scan 
level, followed by scan-length gain calibration solutions that were applied to 
each target individually.  Once the data were passband and phase-calibrated in 
this way they were also fully reduced within CASA owing to the point-like nature 
of the  background targets.  For each polarization and baseband
the final spectra were extracted as the vector phase average over all visibilities,
without gridding or mapping the data.  The vector average spectra were produced in 
CASA's plotms visualizer for uv-visibilities and exported to standard 
singledish software where the polarizations were co-added and very small linear 
baselines amounting typically to 0.01\% of the continuum were removed from 
end to end across each of the basebands.  

The optical depth-column density conversion factors in Table 1 were computed 
assuming rotational excitation in equilibrium with the cosmic microwave 
background.  The validity of this assumption is verified by the excellent
agreement between column densities of \cyclic\ derived here and previously
by \cite{LucLis00C2H} from observations of a higher-frequency transition
arising out of the same lower level (see Table 3).  Note that only
one spin-ladder of either version of  C$_3$H$_2$ was actually observed here,
ortho-\cyclic\ and para-\linear, and the tabulated conversion factors
apply only to the ortho or para version that was observed.  Nominally, 
the ortho:para ratio is 3:1 in either species so that N(\cyclic) = 4/3 
N(o-\cyclic), N(\linear) = 4 N(p-\linear) and N(\linear)/N(\cyclic) 
= 3 N(p-\linear)/N(o-\cyclic).  3:1 ortho:para ratios are used
to convert from specific to total molecular column density in Tables
3 and 4.  The mean $o$-\cyclic/$p$-\cyclic\ ratio measured by 
\cite{LucLis00C2H} in three directions (including B0355 and B0415)
was 3.0, with considerable scatter. 

Table 3 gives total molecular column densities per unit reddening using
the integrated optical depths shown in Table 2 and 3:1 ortho:para ratios,
and Table 4 compares fractional abundances in diffuse
clouds with those toward TMC-1, using mean quantities and upper limits 
from Table 3 for the diffuse gas and various results taken from the literature 
including N(\HH) $=1\times10^{22}~\HH\pcc$ \citep{OhiIrv+92} for TMC-1. 
The fractional abundances of  C$_3$H$_2$, \cfh\ and \anicfh\ in diffuse clouds
in Table 4 were derived from the mean column densities in the last row of Table 3  
and scaled to X(\cch) = $6 \times 10^{-8}$ \citep{LucLis00C2H,GerKaz+11}.

\section{The abundances of small hydrocarbons in diffuse and dark clouds}

\begin{figure*}
\includegraphics[height=13.8cm]{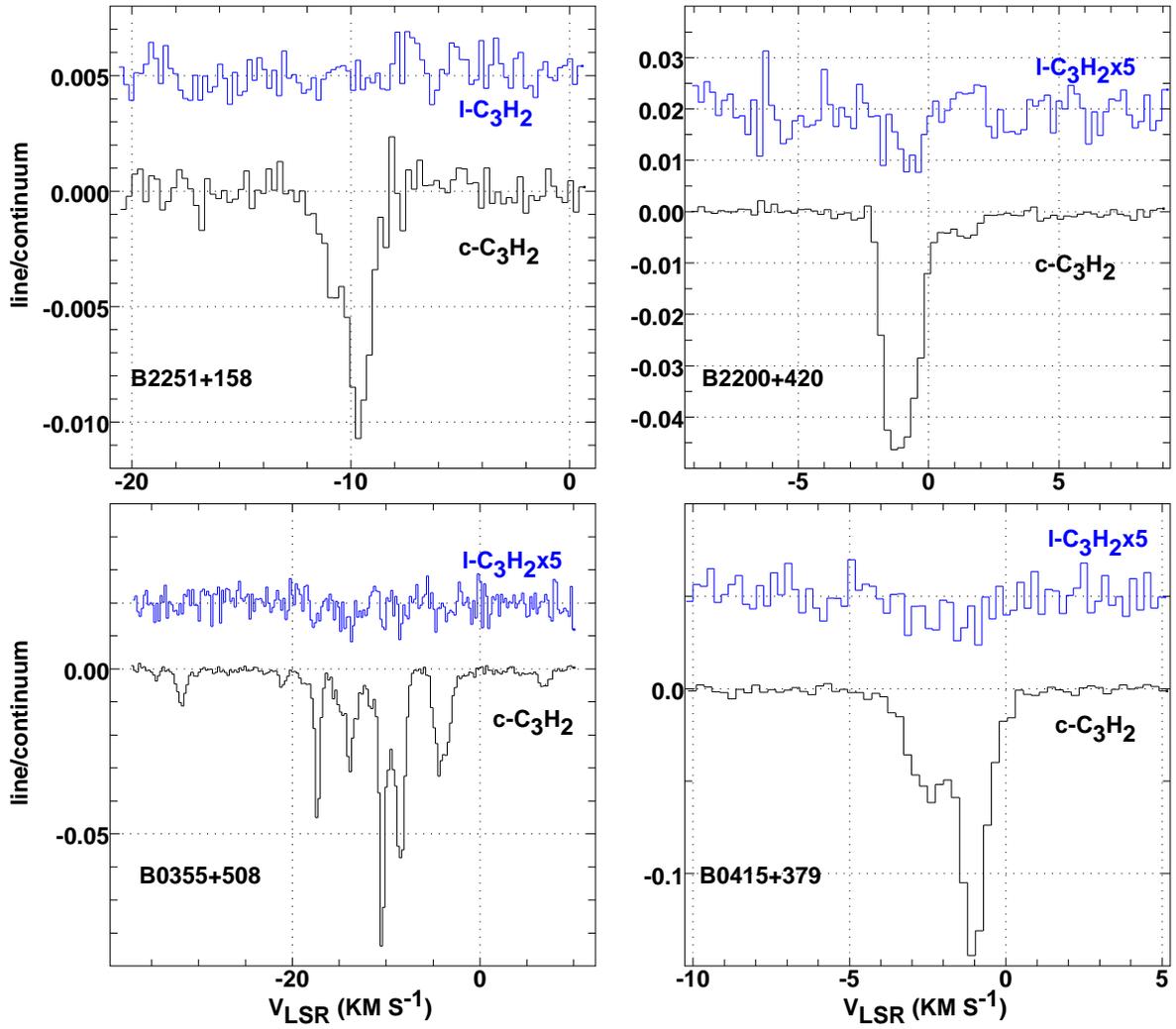}
  \caption[]{Observed absorption line profiles of para linear- and ortho cyclic-
  C$_3$H$_2$ toward four continuum sources.  The profiles of \linear\ have
 been vertically displaced and scaled as indicated. } 
\end{figure*}

Spectra of \cyclic\ and \linear\ are shown in Fig. 1.  From Table 3 we see that 
N(\linear)/N(\cyclic) $\approx 0.06$ and 
N(\linear)/\EBV\ $\approx 2\pm 1 \times10^{11}$ (mag.)$^{-1}$.  The column 
densities of \linear\ in local diffuse clouds are some three orders of magnitude 
smaller than those calculated by \cite{MaiWal+11} in order to explain 
the DIBs (see also \cite{KreGal+11} and \cite{OkaMcC11}).  

Beyond the manifest irrelevance of \linear\ to the DIB problem it is possible
to make some interesting statements about the abundances of small hydrocarbons 
in diffuse clouds.  

\begin{itemize}

\item The ethynyl radical and polyyne (polyyne=C$_n$H) \cch\ is by far the most 
abundant of the small polyatomic hydrocarbons, 
with N(\cch)/\EBV\ = $7\pm 2 \times10^{13}\pcc$/mag.
The abundance of \cch\ relative to \HH\ varies little in diffuse gas
with estimates X(\cch) = $4-7\times10^{-8}$ \citep{LucLis00C2H,GerKaz+11}
and X(\cch) is nearly the same in diffuse gas and toward TMC-1 (see Table 4).
\cch\ is no less abundant than CH in diffuse gas (where X(CH) = 
$4\pm 0.5 \times10^{-8}$ according to \cite{SheRog+08} and 
\cite{WesGal+10}) and is some three times more abundant than CH 
in TMC-1, see Table 4.  Note that this value for X(\cch) implies 
a characteristic value for the molecular hydrogen fraction 
\fH2\ = 2N(\HH)/N(H) = 2N(\HH)/(N(H I)+2N(\HH)) $\approx$ 0.4, 
given that we measured column densities per unit reddening and
N(\cch)/\EBV\ = (N(\cch)/N(\HH))$\times$(N(\HH)/N(H))$\times$(N(H)/\EBV)
with N(H)/\EBV\ = $5.8 \times 10^{21}{\rm H}\pcc$ mag$^{-1}$ 
\citep{BohSav+78,RacSno+09}.  The same value was inferred for the diffuse gas 
as a whole by \cite{LisPet+10}.

\item N(\linear)/\EBV\ varies by a factor 3, somewhat more than
for \cch\ or \cyclic\ (factors $< 2$ in Table 3) but, as with \cch, 
its fractional abundance relative to \HH\ is about the same in diffuse gas 
and toward TMC-1 (Table 4), X(\linear) = $2\times10^{-10}$.  So quite 
generally one has X(\linear)/N(\cch) $\approx 1/350$.  

\item \cyclic\ and \cch\ appear in the fixed proportion N(\cyclic)/N(\cch) 
= 1/21 \citep{LucLis00C2H,GerKaz+11} in diffuse clouds, implying that 
X(\cyclic) = $2-3 \times 10^{-9}$ with little variation.  X(\cyclic) is 
3-4 times smaller in diffuse clouds compared to TMC-1
\footnote{\cite{FosCer+01} mapped emission from both \linear\ and \cyclic\ 
around the position of the polyyne peak in TMC-1.}
 and the ratio N(\linear)/N(\cyclic) =
17 in diffuse clouds is larger for this reason.
 
\item The column density of \cfh\ has previously not been 
strongly-constrained in diffuse clouds (see \cite{LucLis00C2H}), 
but \cfh\ is at most 14\% as abundant as \cch\ in the two directions for 
which we have the best limits.  The fractional abundance of \cfh\ in 
TMC-1 is quite uncertain (Table 4), with estimates ranging between 
$1-9 \times 10^{-8}$ implying 1/6 $\la$ X(\cfh)/X(\cch) $\la$ 1.5.  
All of the \cfh\ column densities  may eventually
require re-scaling because the permanent dipole moment of \cfh\ is 
uncertain (E. Herbst and S. Yamamoto, private communication).  

\item The present results represent the first search for \anicfh\
in diffuse gas.  Despite the high free electron abundance n(e)/n(H) $\ga$ 
n(C\p)/n(H) $= 1.6\times 10^{-4}$ \citep{SofLau+04}, 
the abundance of \anicfh\ is very small, X(\anicfh)/X(\cch) $\la 0.001$
or X(\anicfh) $\la 6\times10^{-11}$.  The abundance of molecular anions is 
most likely limited in diffuse gas because so much of the ambient hydrogen 
is in atomic form and anion neutralization by atomic hydrogen is rapid 
\citep{EicSno+07}.  Our limits on N(\anicfh) are comparable to those reached 
by \cite{AguCer+08} toward TMC-1, which implied X(\anicfh) $< 4\times 10^{-12}$,
but Cordiner (unpublished) recently detected \anicfh\ toward TMC-1 with
X(\anicfh) $= 1 \pm 0.25\times 10^{-12}$.

\end{itemize}

\section{Summary and discussion}

The sightlines in this work substitute distant radio sources for early-type 
stars used in optical/uv absorption-line studies, uniquely
providing access to a diverse set of polar polyatomic molecules.  Nonetheless 
there are comparable column densities of all species that are accessible in
both wavelength regimes, OH, CO and CN \citep{LisLuc96,LisLuc98,LisLuc01} 
and comparable thermal pressures as judged in the radio by the rotational 
excitation of CO \citep{LisLuc98}.  CO column densities are small and
the gas-phase carbon is inferred to reside overwhelmingly in C\p\ even when 
N(C\p) is not measured directly, implying that the ionization fraction is 
similar even if the flux of ionizing photons will occasionally be larger 
in optical absorption line studies.  Although there is no theory that 
reliably predicts the abundances of polyatomic molecules in diffuse clouds, 
similar circumstances should affect both \linear\ and DIBs no matter which
measurement technique is employed

The simple hydrocarbons observed and discussed here generally have 
well-defined abundances with respect to reddening and \HH\ over a 
wide range of conditions, especially CH, \cch\ and \cyclic, with slightly 
larger variations for \linear.  Bulk effects naturally introduce some 
degree of superficial correlation with \EBV\ even for unrelated species that 
are mixed into the ISM along the line of sight, but proportionality between the
column densities of  small hydrocarbons and N(\HH) reflects a concentration
within regions of high molecular fraction, typically \fH2 $\approx$ 0.4 as noted 
above.  This may make the small hydrocarbons rather poorly suited to be
candidate carriers for the most heavily-studied DIBs \citep{FriYor+11,VosCox+11},
which do not have obvious associations with N(\HH) and are more usually
attributed to regions of low molecular fraction.

In any case, the column density per unit reddening of \linear\ in diffuse clouds 
is a factor 2000 below the value 
N(\linear)/\EBV\ $= 4 \times 10^{14}~\pcc$ mag$^{-1}$ that was 
hypothesized by \cite{MaiCha+11} in order that \linear\ could be 
the carrier of the DIBs at 4881\AA\ and 5450\AA.  The most abundant
hydrocarbons in diffuse gas, as toward TMC-1, are CH and \cch\ with 
X(CH) = $4 \times 10^{-8}$ and X(\cch) $= 6\times 10^{-8}$.
Taken relative to \cch\ in diffuse gas one has X(\cyclic)/X(\cch) = 
1/21 and X(\linear)/X(\cch) = 1/350 from the mean
values given in Table 3, with X(\cfh)/X(\cch) $< 1/7$ in the two best cases 
and X(\anicfh)/X(\cch) $ \la 1/1000$ quite generally.
 
\cite{GreCar+11} review the history of (failed) DIB attributions, including
\linear\ and the equally recent example of H-\cfh-H\p, proposed by 
\cite{KreBel+10} and criticized by \cite{MaiCha+11}.  Other attributions
recently deemed to be insupportable include heavier species such as
the simplest PAH naphthalene (C$_{10}$H$_8$), anthracene (C$_{14}$H$_{10}$) and 
their cations \citep{GalLee+11}.  Perhaps still hanging in the balance, but by
no means generally accepted, is a tentative attribution to C\HH CN$^-$\ by 
\cite{CorSar07}.

Species as complicated as naphthalene are likely to prove as elusive at radio
wavelengths as in the optical/uv domain, but the sort of work performed here
may be straightforwardly extended to other polar species such as C$_3$H and C\HH CN,
especially considering the modest observing times required and the increasing 
availability of wider instantaneous bandwidths at the GBT and VLA.  Searches 
may also be possible for non-polar species that are observable in their magnetic 
dipole transitions as recently discussed by \cite{MorMai11}.

\acknowledgments

  The National Radio Astronomy Observatory is operated by Associated
  Universities, Inc. under a contract with the National Science Foundation.
  HL and MG were partially funded by the grant ANR-09-BLAN-0231-01 from the 
  French {\it Agence Nationale de la Recherche} as part of the SCHISM 
  project (http://schism.ens.fr/).  MAC acknowledges support from the 
  NASA Astrobiology Institute through the Goddard Center for Astrobiology.

{\it Facilities:} \facility{Jansky VLA}




\bibliographystyle{apj}




\begin{table}
\caption[]{Species and transitions observed and column density-optical 
depth conversion factors}
{
\small
\begin{tabular}{lccccc}
\hline
Species & ortho/para & transition & frequency& log(A$_{kj}~\ps)^a$ & N(X)/$\int\tau dv^b$ \\
        &   &                        & MHz  &     & $\pcc$ (\kms)$^{-1}$ \\
\hline
\anicfh &   & 2-1             & 18619.76 & -5.938& $5.41\times10^{12} $ \\
\cfh    &   & N=2-1 J=5/2-3/2 F=2-1  & 19014.72 &-7.671 & $1.14\times10^{15} $ \\
\cfh    &   & N=2-1 J=5/2-3/2 F=3-2  & 19015.14 & -7.616& $7.12\times10^{14} $ \\
\cyclic & o & 1$_{10}$-1$_{01}$      & 18343.14 & -6.374  & $1.35\times10^{13} $ \\
\linear & p &  1$_{01}$-$0_{00}$     & 20792.59 & -6.232 & $9.00\times10^{12} $ \\
\tableline
\end{tabular}}
\\
$^a$ www.splatalogue.net \\
$^b$ for the observed ortho or para version only, assuming rotational \\
excitation in equilibrium with the cosmic microwave background \\
\end{table}

\begin{table*}
\caption[]{Continuum target, line of sight and line profile  properties}
{
\small
\begin{tabular}{lcccccccc}
\hline
Target &  l  &  b & \EBV$^a$ & flux$^b$ 
 & EW(o-\cyclic)$^c$ & EW(p-\linear) & EW(\cfh)$^d$ & EW(\anicfh) \\
       &  \degr  & \degr  & mag  & \% 
       & m \ps &  m \ps & m \ps  & m \ps \\

\hline
B0355+508 & 150.38 &-1.60 & 1.50 & 34 
 & 320$\pm$2  & 4.0$\pm$1.4  & $<$16.8$^e$ & $<$13.3 \\
B0415+379 & 161.67 & -8.82 & 1.65 & 9 
 & 250$\pm$3 & 7.8$\pm$1.8  & $<$40.2 & $<$ 11.6 \\
B2200+420 & 92.59 & -10.44 & 0.33 & 18 
 & 84$\pm$1 & 2.8$\pm$0.7 & $<$7.40& $<$ 6.2 \\
B2251+158 & 86.11 & -38.18 &0.11 & 51 
 & 170$\pm$1 & $<$ 2.4  & $<$4.8 & $<$ 4.6 \\ 
\tableline
\end{tabular}}
\\
$^a$from \cite{SchFin+98} \\
$^b$ relative to 3C84 (S$_\nu \approx 16$ Jy)\\
$^c$ EW=$\int\tau dv$ for the observed transition, for \cfh\ applicable to either transition 
observed  \\
$^e$ all upper limits are $3\sigma$
\\
\end{table*}

\begin{table}
\caption[]{Total molecular column densities per unit reddening$^a$}
{
\small
\begin{tabular}{lcccccc}
\hline
Target & \cyclic$^b$ & \linear$^c$ & \cfh & \anicfh & \cch$^d$ & \cyclic$^d$ \\
\hline
B0355 & 38.5$\pm0.20$ & 0.96$\pm0.34$ & $<$ 66E$^e$ & $<$ 0.48 & 607 & 41.4\\
B0415 & 28.2$\pm0.30$ & 1.76$\pm0.39$  & $<$ 150 & $<$ 0.39 & 519 & 25.9\\
B2200 & 45.7$\pm0.48$ & 3.05$\pm0.76$  & $<$ 130 & $<$ 1.00 & 939 & 43.5\\
B2251 & 21.1$\pm1.45$ & $<$ 7.60 & $<$ 260 & $<$ 2.32 & 613 & \\
Mean & 33$\pm11$ & 1.9$\pm1.1$ &  &  &670$\pm180$ & \\ 
\tableline
\end{tabular}}
\\
$^a$ Units are $10^{11} \pcc$ mag$^{-1}$. \\
$^b$N(\cyclic)= $(4/3)\times$N($o$-\cyclic) \\
$^c$N(\linear)= $4\times$N($p$-\linear) \\
$^d$ N(\cch) and N(\linear) from \cite{LucLis00} \\
$^e$ all upper limits are $3\sigma$ \\
\end{table}

\begin{table*}
\caption[]{Fractional molecular abundances in diffuse clouds$^a$ and TMC-1$^{a,b}$}
{
\small
\begin{tabular}{lcccccc}
\hline
Target & X(\cyclic) & X(\linear) & X(\cfh) & X(\anicfh) & X(\cch) & X(CH) \\
\hline
Diffuse & 3 & 0.2 & $<$8 $^{c,d}$ & $<$0.07$^{c,e}$ & 60  & 40 \\ 
TMC-1 & 10$^{f,g}$& 0.3$^{h}$ & 20$^{f,i}$ & 0.001$^j$ & 60$^{f,k}$ & 20$^{f,g}$\\
\tableline
\end{tabular}}
\\
$^a$ In units of $10^{-9}$. 
Results from Table 3 and X(CH) from \cite{SheRog+08} and \cite{WesGal+10} \\
$^b$ N(\HH) = $10^{22}$ for TMC-1 \citep{OhiIrv+92} \\
$^c$ all upper limits are $3\sigma$ \\
$^d$ X(\cfh)/X(\cch) $<$ 0.14 for B0355 and B2200 \\
$^e$ X(\anicfh)/X(\cch) $<$ 0.0011 for B0355, B0415 and B2200 \\
$^f$ \cite{OhiIrv+92} \\
$^g$ \cite{SmiHer+04}; \cite{TurHer+00} \\
$^h$ \cite{CerGot+91}, 4E-10 from \cite{TurHer+00}. \\
$^i$ $10\times10^{-9}$ from \cite{TurHer+00}; $70\times10^{-9}$ from \cite{AguCer+08};
 $90\times10^{-9}$  from \cite{SmiHer+04} \\
$^j$ Cordiner (2012) unpublished.  \cite{AguCer+08} gave X(\anicfh) $< 0.004\times10^{-9}$ \\
$^k$ \cite{SakSar+10}. \cite{SmiHer+04} and \cite{TurHer+00} give 2E-8 \\
\\
\end{table*}

\end{document}